\documentclass[a4paper,11pt]{article}

\usepackage{jinstpub} 

\usepackage{subcaption} 

\usepackage{lineno}

\title{\boldmath  Multichannel read-out for arrays of metallic magnetic calorimeters}

\author[1,*]{F. Mantegazzini}
\author[1,*]{S. Allgeier}
\author[1]{A. Barth}
\author[1]{C. Enss}
\author[1]{A. Ferring-Siebert}
\author[1]{A. Fleischmann}
\author[1]{L. Gastaldo}
\author[1]{R. Hammann}
\author[1]{D. Hengstler}
\author[1,2]{S. Kempf}
\author[1]{D. Richter}
\author[1]{D. Schulz}
\author[1]{D. Unger}
\author[1]{C. Velte}
\author[1,2]{M. Wegner}

\affiliation[1]{\textit{\normalsize Kirchhoff-Institute for Physics, Im Neuenheimer Feld 227, 69120 Heidelberg, Germany.}}
\affiliation[2]{\textit{\normalsize Institute of Micro- and Nanoelectronic Systems, Karlsruhe Institute of Technology, Hertzstraße 16, 76187 Karlsruhe, Germany.}}
\affiliation[*]{\textit{\normalsize Corresponding authors}}

\emailAdd{federica.mantegazzini@kip.uni-heidelberg.de}
\emailAdd{steffen.allgeier@kip.uni-heidelberg.de}

\abstract{Metallic magnetic micro-calorimeters (MMCs) operated at millikelvin temperature offer the possibility to achieve eV-scale energy resolution with high stopping power for \mbox{X-rays} and massive particles in an energy range up to several tens of keV. This motivates their use in a wide range of applications in fields as particle physics, atomic and molecular physics. Present detector systems consist of MMC arrays read out by 32 two-stage SQUID read-out channels. In contrast to the design of the detector array and consequently the design of the front-end SQUIDs, which need to be optimised for the physics case and the particles to be detected in a given experiment, the read-out chain can be standardised. 
We present our new standardised 32-channel parallel read-out for the operation of MMC arrays to be operated in a dilution refrigerator. The read-out system consists of a detector module, whose design depends on the particular application, an amplifier module, ribbon cables from room temperature to the millikelvin platform and a data acquisition system. In particular, we describe the realisation of the read-out system prepared for the ECHo-1k experiment for the operation of two 64-pixel arrays. 

The same read-out concept is also used for the maXs detector systems, developed for the study of the de-excitation of highly charged heavy ions by X-rays, as well as for the MOCCA system, developed for the energy and position sensitive detection of neutral molecular fragments for the study of fragmentation when molecular ions recombine with electrons.
The choice of standard modular components for the operation of 32-channel MMC arrays offer the flexibility to upgrade detector modules without the need of any changes in the read-out system and the possibility to individually exchange parts in case of damages or failures. }


\arxivnumber{2102.1110} 

\begin{document}
\maketitle
\flushbottom

\section{Detector set-up}

In the following a close look to the ECHo detector is presented, as an example of the variety of detectors.
The detector chip for the first phase of the ECHo experiment, \mbox{ECHo-1k}, consists of an MMC array of 36 SQUID read-out channels \cite{ECHo-1k}.
Two of the channels on the chip are devoted to test purposes, namely detector diagnostics and chara\-cte\-risation, two channels are dedicated to temperature monitoring and the remaining 32 channels can be used for measuring the $^{163}$Ho spectrum after the source is embedded via ion-implantation \cite{Gastaldo_NIMA}.
Each SQUID channel reads out two meander-shaped pick-up coils \cite{Fle2005}, as described above, arranged in a gradiometric configuration allowing for the identification of the pixel where a particle event occurred via the polarity of the signal.
Moreover, the gradiometric setup makes the MMC channel fairly insensitive to fluctuations of the substrate temperature, as the temperature variations happen simultaneously in both pixels.
In order to operate the detector, a persistent current is injected in the meander loop by means of a dedicated persistent current switch, which is used to locally break the superconductivity of the closed niobium loop formed by the two pick-up coils \cite{ECHo-1k}.

\begin{figure}[h!]
    \centering
    \includegraphics[width=0.7\linewidth]{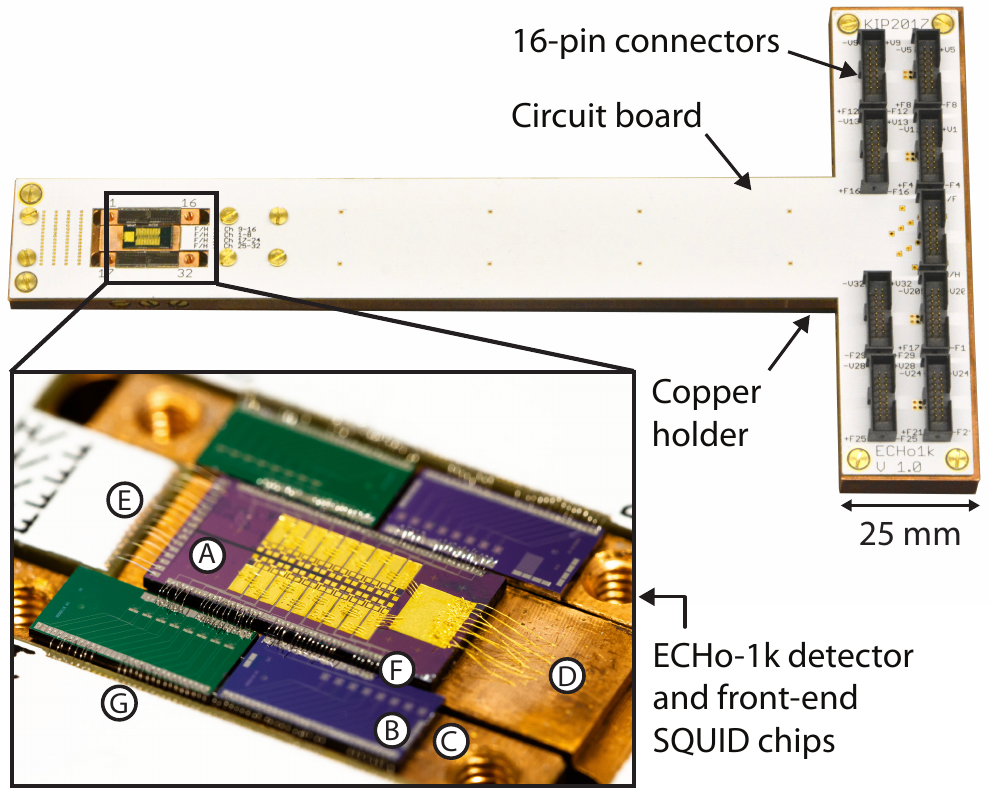}
    \caption{ECHo-1k detector sample holder made of copper, mostly covered by a circuit board with white solder mask. In the inset a close-up view on the ECHo-1k detector chip (A) glued onto the holder is shown. The detector chip is wire-bonded to four front-end SQUID chips (B), each one containing 8 SQUID channels, which are glued on separate copper blocks (C). The gold wire-bonds connecting the on-chip thermal baths with the copper holder (D) as well as the aluminium wire-bonds between detector chip and circuit board (E), between detector chip and front-end SQUID chips (F) and between front-end SQUID chips and circuit board (G) are visible.
    }
    \label{FIG:det_setup}
\end{figure}

The chip size is $10 \, \mathrm{mm} \times 5 \, \mathrm{mm}$ and it is designed to be wire-bonded to external front-end SQUID read-out chips.
The SQUID chips are fabricated in house\footnote{Cleanroom facility at the Kirchhoff-Institute for Physics, Heidelberg University.} employing niobium as superconducting material and aluminium oxide as barrier material within the Josephon junctions.
The detector operational temperature is about $20 \, \mathrm{mK}$ and is reached by mounting the detector set-up at the mixing chamber plate of a dilution refrigerator.
In order to ensure optimal thermalisation, the detector chip is glued on a custom-built T-shaped holder made of annealed OFHC\footnote{Oxygen Free High Conductivity.} copper with a size of $2.7 \, \mathrm{cm} \times 19.3 \, \mathrm{cm} \times 0.8 \, \mathrm{cm}$, as shown in figure \ref{FIG:det_setup}.
Typically two glues are used to assemble the set-up: Pritt-Stick\footnote{Produced by Henkel AG \& Co. KGaA.} and GE varnish 7031.
Additionally, the on-chip gold thermal baths are connected via gold wire-bonds to the copper holder for a better thermalisation of the detector chip (figure \ref{FIG:det_setup} inset, A).

The front-end SQUID chips are glued onto two separate copper blocks that can be inserted and screwed to the main holder right next to the detector area. This makes the preparation of the detector set-up more flexible, as the SQUID blocks can be easily exchanged allowing for a modular approach, where possibly faulty components can be replaced. The SQUID chips are electrically connected to the detector via aluminium wire-bonds bridging the slits that separate the SQUID copper blocks from the detector holder (figure \ref{FIG:det_setup}, F). The SQUID chips are wire-bonded to the detector on one side and on the other side to a customised double layer circuit board (figure \ref{FIG:det_setup}, G), which is fixed on top of the copper holder. The bond-pads at the bottom periphery of the detector are dedicated to the persistent current injection and they are directly wire-bonded to copper lines on the circuit board (figure \ref{FIG:det_setup}, E).

\begin{figure}[h!]
    \centering
    \includegraphics[width=0.62\linewidth]{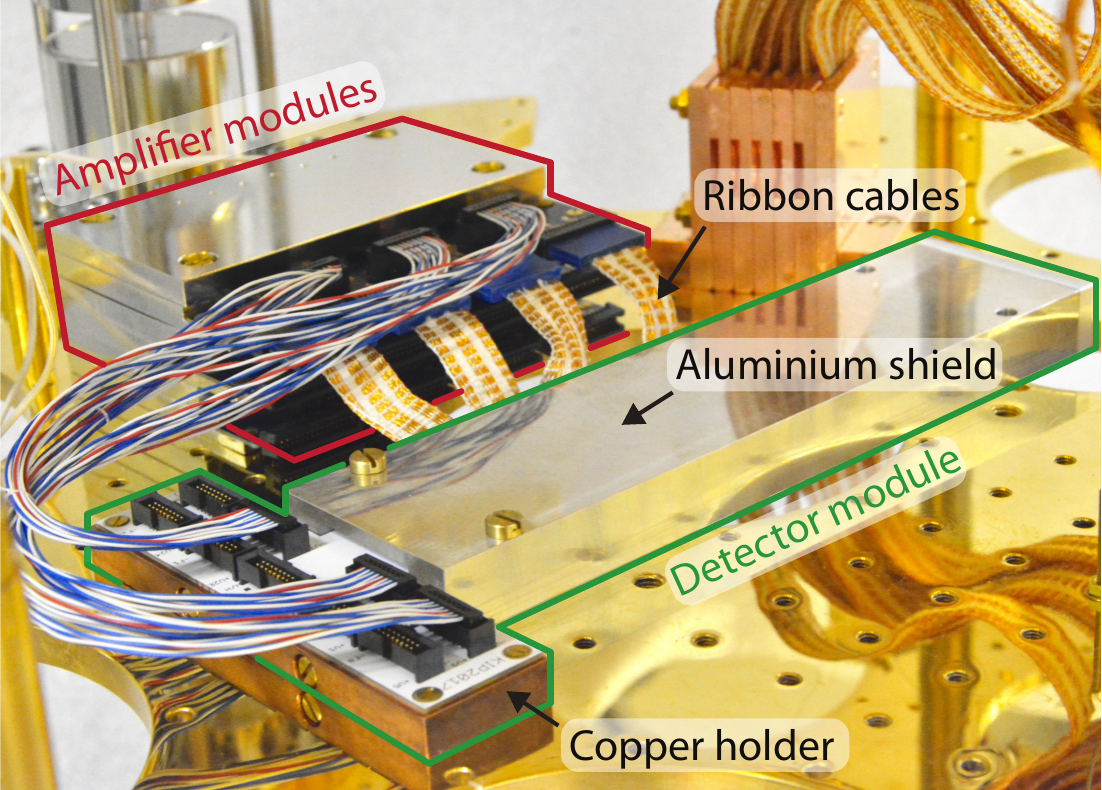}
    \caption{The complete cold part of the ECHo-1k set-up placed on the cryostat mixing chamber plate. It includes the shielded detector module and the amplifier stage.
    }
    \label{FIG:complete_setup}
\end{figure}

The T-shaped detector holder can be inserted in an aluminium shield, as shown in figure \ref{FIG:complete_setup}. The shield becomes superconducting for temperatures below $1.2 \, \mathrm{K}$ and therefore, due to the Mei\ss ner–Ochsenfeld effect, it shields detector and front-end SQUIDs against changes of external magnetic fields. The ratio between the depth and the longer side length of the opening is about 8.6 and ensures a high shielding level.

The circuit board is equipped with nine double-row 16-pin connectors\footnote{Connector type SHF-108-01-L-D-TH produced by Samtec, 520 Park East Boulevard, New Albany, IN 47150, USA.}: eight of them are connected to the SQUID lines and one is connected to the lines dedicated to the injection of the persistent current in the pick-up coils. 
Four wires are necessary to read-out each front-end SQUID channel and therefore each SQUID connector corresponds to four front-end SQUID channels. The front-end SQUID output signals are transferred to amplifier SQUIDs via tin-coated copper wires which connect the SQUID connectors on the detector circuit board to the amplifier module, as visible in figure \ref{FIG:complete_setup}. The wires are commercially available with different wire counts in three different lengths\footnote{Produced by Samtec.}.
The second stages consist of 16 SQUID-series arrays and acts as a low temperature amplifier stage.
The working principle of a two-stage SQUID read-out and the custom-built amplification module will be described in the following section.

\section{Two-stage SQUID read-out}
\label{SEC:SQUID_readout}

\begin{figure}[h!] 
	\centering
	\includegraphics[width=0.8\textwidth]{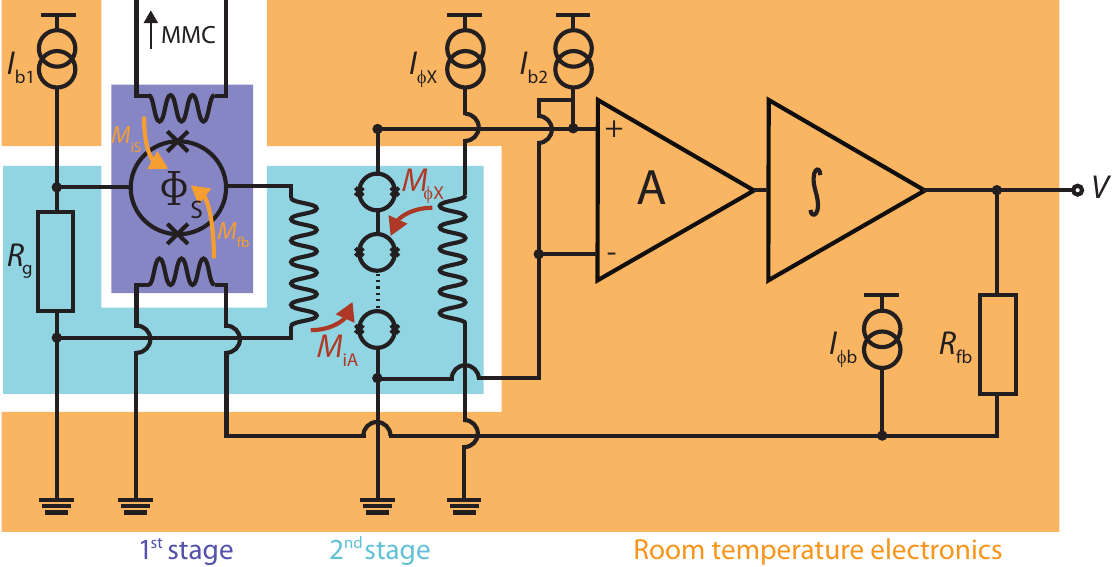}
	\caption{Schematic drawing of a two-stage SQUID configuration, where the first stage is marked in violet, the second stage is marked in blue and the room temperature electronics is marked in orange.} 
	\label{FIG:two_stage}
\end{figure}

The scheme of a two-stage dc-SQUID configuration is depicted in figure \ref{FIG:two_stage}. The signal from the MMC detector is coupled to a front-end dc-SQUID, the output of which is coupled into the input of a second-stage SQUID array (i.e. 16 dc-SQUIDs in series) that serves as a low noise amplifier \cite{Kem15a}. 
The SQUID response is linearised by a flux-locked loop (FLL) feedback mechanism applied to keep the front-end SQUID at a constant working point \cite{SQUID_handbook}.
In the two-stage set-up, the front-end SQUID is operated in voltage bias mode by choosing the load resistor $R_\mathrm{g}$ much smaller than the dynamical resistance of the front-end SQUID, to decrease the power dissipation of the SQUID chip and, in turn, of the detector chip. The second-stage amplifier SQUID array is operated in current bias mode. The design and fabrication of front-end and amplifier SQUIDs are tailored to the specific requirements for MMC detectors\footnote{The SQUID chips are designed and fabricated in house at the Kirchhoff-Institute for Physics, Heidelberg University.} \cite{Kem15a}. In particular, the input coil inductance of the front-end SQUIDs is designed to match the inductance of the detector meander-shaped pick-up coils.

\begin{figure} [h!] 
\centering
\begin{subfigure}{.45\textwidth}
  \centering
  \includegraphics[width=0.85\linewidth]{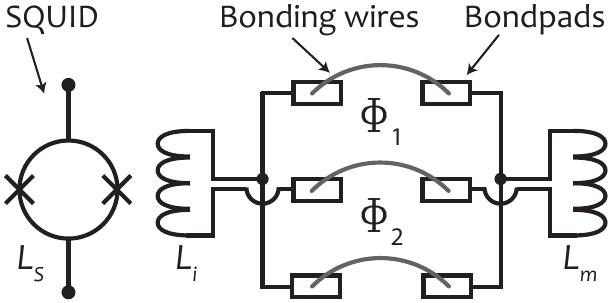}
  \caption{}
  \label{SUBFIG:3-bond_schema}
\end{subfigure}%
\begin{subfigure}{.55\textwidth}
  \centering
  \includegraphics[width=\linewidth]{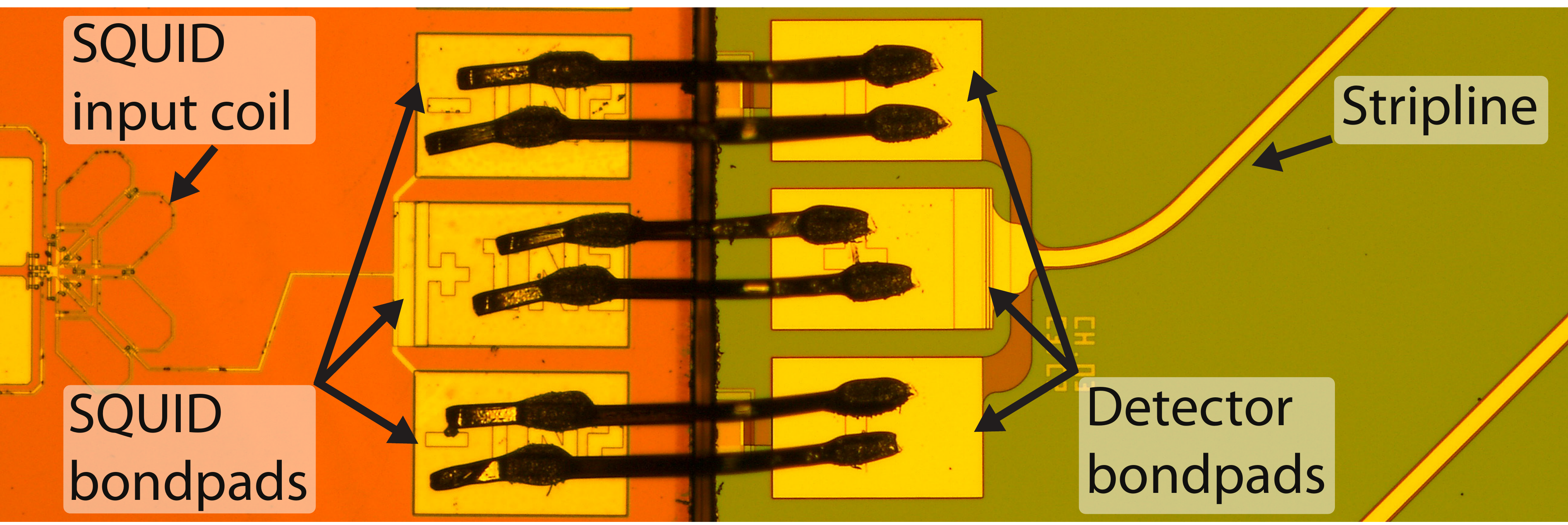}
  \caption{}
  \label{SUBFIG:3-bond_schema_microscope}
\end{subfigure}
\caption{a) Schematic sketch of the optimised three-bond scheme which allows for reduced pick-up of external magnetic fields. The magnetic fluxes $\Phi_1$ and $\Phi_2$ are generating currents with the same magnitude but with opposite polarity, leading to a first gradiometer layout and therefore to the cancellation of additional magnetic interference. b) Microscope picture of double aluminium wire bonds that connect SQUID and detector using the optimised three-bond scheme.}
\label{FIG:3-bond_scheme}
\end{figure}

The pick-up coils of the detector and the input coil of the SQUID are connected by superconducting aluminium wires to form a completely superconducting circuit. Changes in external magnetic fields or vibrations of the bonding wires can generate currents in this loop that lead to an increased noise level. An optimised bonding scheme where on both chips, SQUID and detector, one of the two bond pads is duplicated has been developed, as shown in figure \ref{FIG:3-bond_scheme}. When the resulting three pairs of bond pads are connected with bonding wires, the arrangement of these wires represents a gradiometer of first order for the pick-up of external magnetic fields. This layout can reduce the additional noise by nearly one to two orders of magnitude.

\begin{figure}[ht]
    \centering
    \begin{subfigure}[b]{0.48\textwidth}
        \centering
        \includegraphics[width=0.95\linewidth]{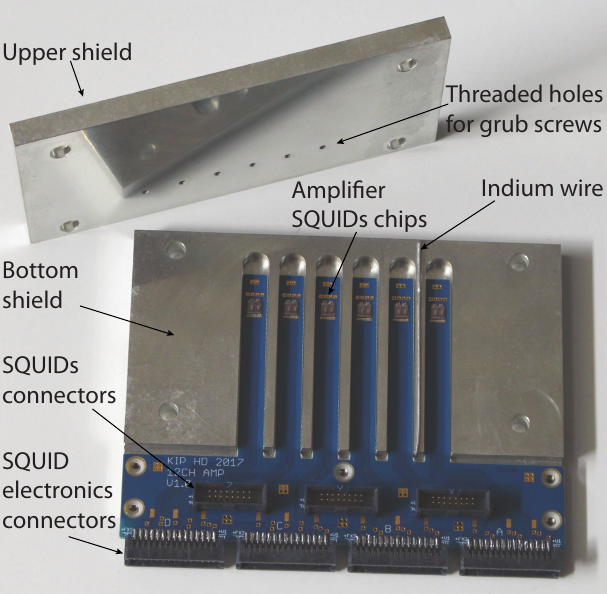}
        \caption{} \label{SUBFIG:array_module_labels}
    \end{subfigure}
    \hfill
    \begin{subfigure}[b]{0.48\textwidth}
        \centering
        \includegraphics[width=0.95\linewidth]{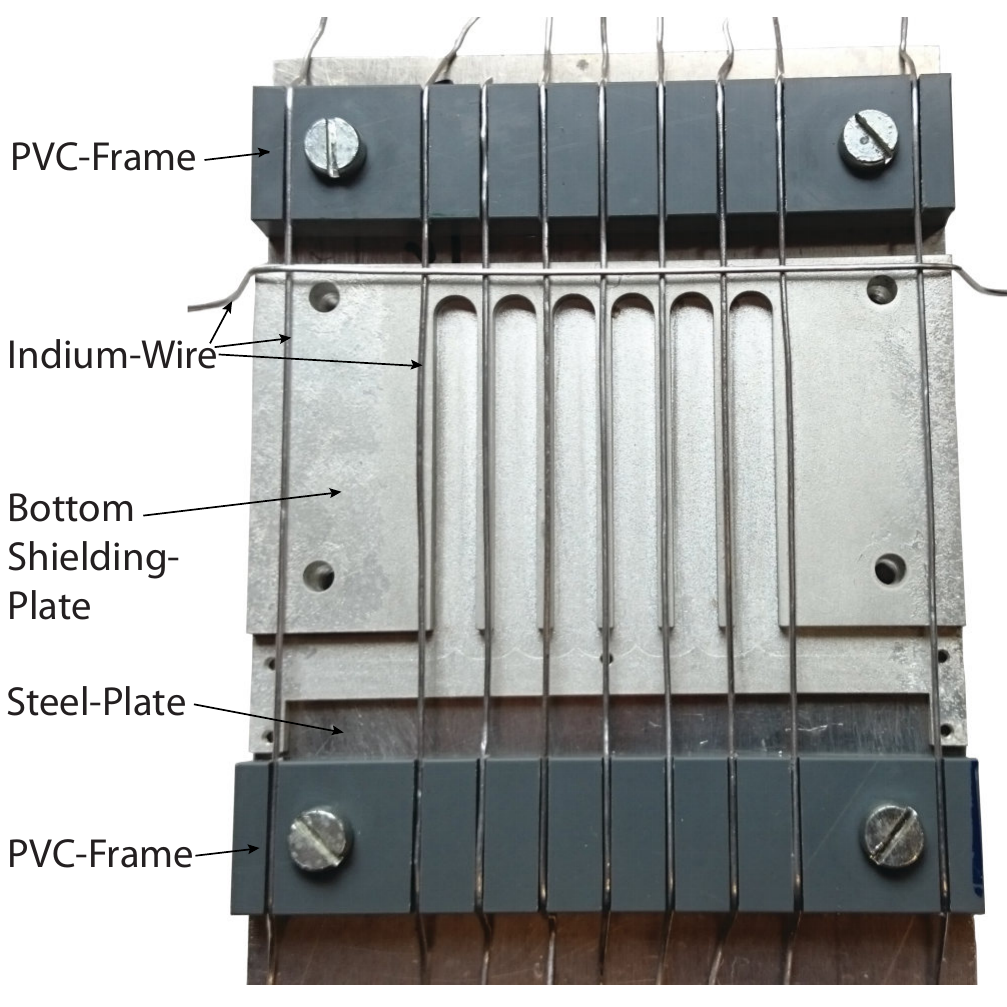}
        \caption{} \label{SUBFIG:array_module_build}
    \end{subfigure}
    \caption{a) A single array module consisting of the custom made circuit board inserted in the bottom half of the tin plated copper shield. The upper part is placed on top to close the module.  b) Set-up for the fabrication of the superconducting shield. The indium wires are positioned on the bottom plate before pressing the second plate on top of it and proceed with the diffusion welding of the two parts.}

    \label{FIG:array_module}
\end{figure}

While the front-end SQUID chips are glued onto the detector holder and wire-bonded to the detector chip, the second-stage amplifier SQUIDs are placed on separate custom-designed modules to reduce power dissipation to the detector module, operated at about $\mathrm{20 \, mK}$. 
The modules are based on the custom-made circuit boards\footnote{Produced by Multi Leiterplatten GmbH, Brunnthaler Straße 2, D-85649 Brunnthal - Hofolding, Germany.} shown in figure \ref{FIG:array_module}.
Each circuit board is equipped with:

\begin{itemize}
    \item six dc-SQUID array chips with a size of $3 \, \mathrm{mm} \times 3 \, \mathrm{mm}$, each hosting two series arrays of dc-SQUIDs, glued to the circuit board at the end of a $45 \, \mathrm{mm}$ long circuit board finger and wire-bonded to the copper leads;
    \item three double row 16-pin connectors\footnote{Connector type SHF-108-01-L-D-TH produced by Samtec.} connecting to the front-end SQUIDs shown in figure \ref{SUBFIG:array_module_labels};
    \item four double row 30-pin connectors\footnote{Connector type TFM-115-01-S-D produced by Samtec.} to which the ribbon type cables (described in section \ref{SEC:cryo}) that connect the amplifier SQUIDs to the room temperature read-out stage are plugged.
\end{itemize}

The circuit board is inserted into a tin-plated copper case.
This case consists of two copper plates that are coated with about $8 \, \mathrm{\upmu m}$ thick layer of tin. Due to its high thermal conductivity, copper ensures a reliable thermalisation of the heat produced by the amplifier SQUIDs. On the other hand, tin enters the superconducting regime at $3.7 \, \mathrm{K}$ and, because of the Mei\ss ner–Ochsenfeld effect, provides a shielding against fluctuating external magnetic fields.
The bottom plate is equipped with $3 \, \mathrm{mm}$ deep millings in which the fingers of the circuit board can be inserted. The bottom plate and the upper plate need to be connected forming a long superconducting cup for each amplifier SQUID. In this way, each chip is placed in a separate superconducting compartment and therefore optimally shielded against external magnetic field fluctuations. 
To reach this goal, the two tinned copper plates are diffusion welded with indium. Indium enters the superconducting regime at $3.4 \, \mathrm{K}$ and it forms an eutectic with tin at about $120 \, ^{\circ} \mathrm{C}$. Pure indium melts at $157 \, ^{\circ} \mathrm{C}$ and pure tin at $232 \, ^{\circ} \mathrm{C}$.
To prepare the diffusion welding both plates are pressed to each other with indium wires of $1 \, \mathrm{mm}$ diameter placed in between them. The indium wires are positioned as shown in figure \ref{SUBFIG:array_module_build}. 
In pressed state the plates are baked for three hours at $140 \, ^{\circ} \mathrm{C}$ in air at normal pressure.

The circuit board is attached to the module case with five screws at the connector side. Therefore, the six fingers of the circuit board can potentially be subjected to vibrations. To prevent that, they are additionally fixed to the module by means of tinned M 1.6 grub screws (also known as set screws) inserted into the threads in the upper part of the module, close to the amplifier SQUID chips.

Three amplifier modules, corresponding to 36 read-out channels, can be stacked forming a compact tower, which can be mounted on the mixing chamber plate of a dilution refrigerator, as depicted in figure \ref{SUBFIG:ECHo_cryo}. The amplifier modules are additionally surrounded by a cryoperm shield \footnote{Produced by Magnetic Shields Ltd, Headcorn Rd, Staplehurst TN12 0DS, UK.}. This soft magnetic shield allows for a reduced background magnetic field during the transition of the tin parts to the superconducting state.

\newpage
\section{Wiring from room temperature to millikelvin} 
\label{SEC:cryo}

The ECHo detectors are operated in a dedicated dry dilution refrigerator of type BF-XLD\footnote{Produced by BlueFors Cryogenics Oy, Arinatie 10, 00370 Helsinki, Finland.} that reaches a base temperature below $7 \, \mathrm{mK}$. This cryostat has a cooling power of $20 \, \mathrm{\upmu W}$ at $20 \, \mathrm{mK}$ at the mixing chamber plate. Along with the capability to host a net load of about $200 \, \mathrm{kg}$, the large available experimental space - corresponding to a cylindrical volume with a diameter of $50 \, \mathrm{cm}$ and a height of $50 \, \mathrm{cm}$ below the mixing chamber plate - and the continuous operation, the cryostat is suitable for the current and future phases of the ECHo experiment. 

\begin{figure}[!htb]
 \centering
 \begin{subfigure}[b]{0.505\textwidth}
 \includegraphics[width=\linewidth]{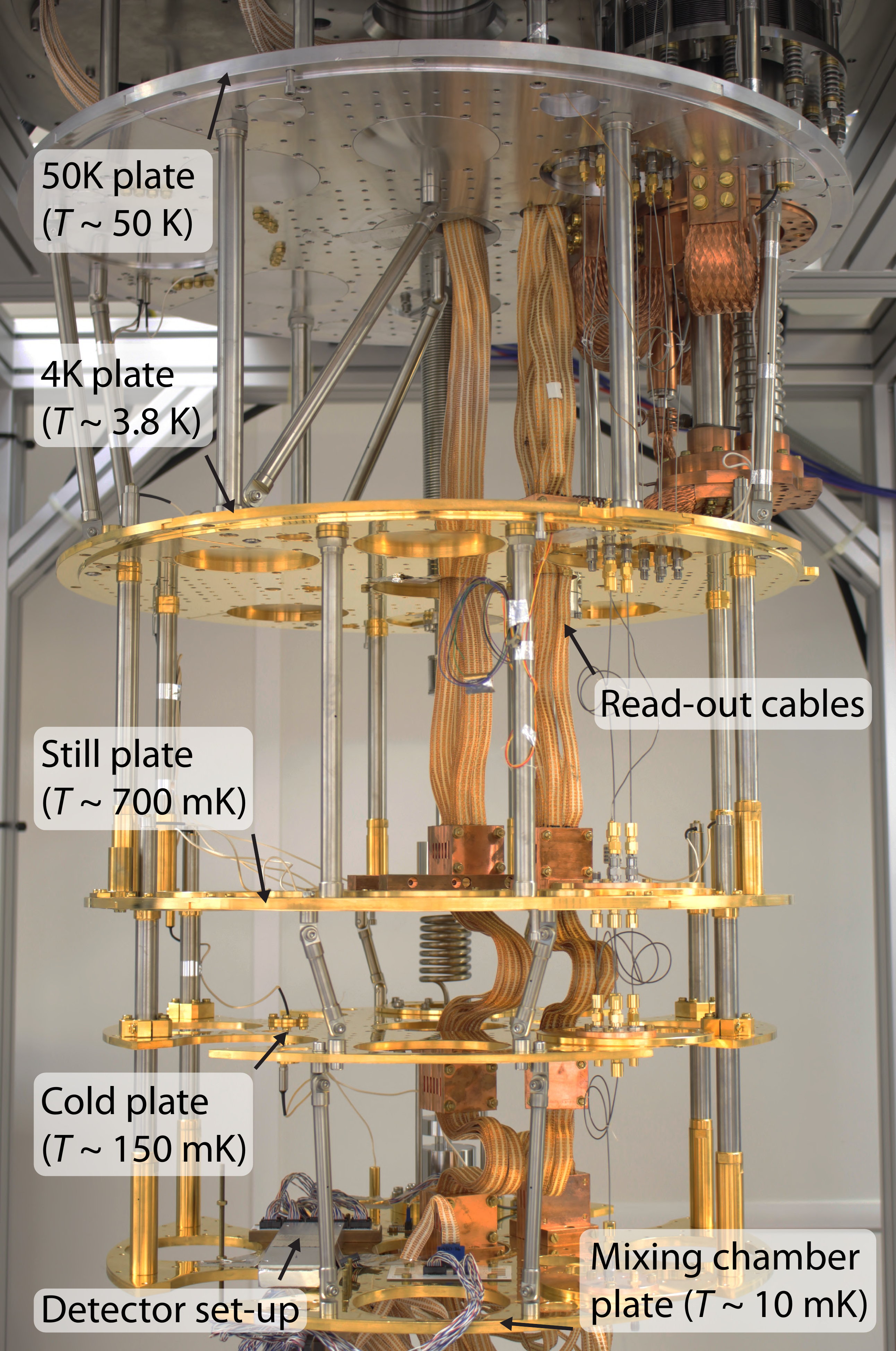}
  \caption{}\label{SUBFIG:ECHo_cryo}
 \end{subfigure}
 \hfill
 \begin{minipage}[b]{0.43\textwidth}
  \begin{subfigure}[b]{\linewidth}
  \includegraphics[width=\linewidth]{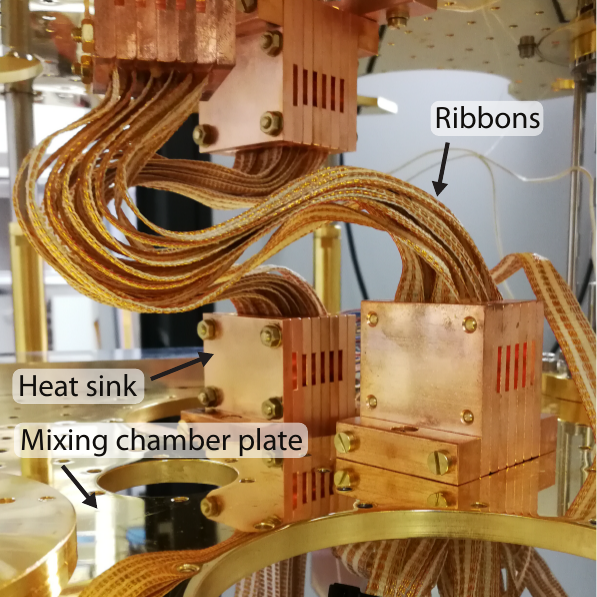}
     \caption{}\label{SUBFIG:heat_sinks}
  \end{subfigure}\\[\baselineskip]
  \begin{subfigure}[b]{\linewidth}
  \includegraphics[width=\linewidth]{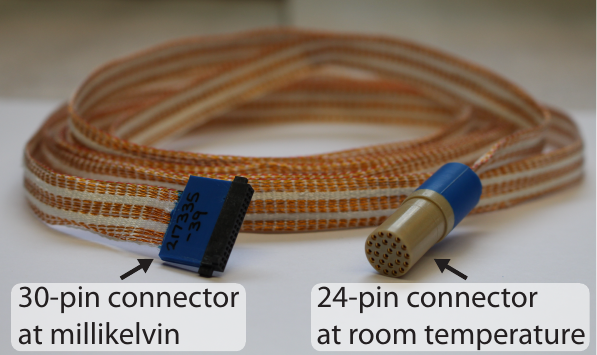}
     \caption{}\label{SUBFIG:ribbon}
  \end{subfigure}
 \end{minipage}
 \caption{a) Open dilution refrigerator dedicated to the ECHo experiment. The temperature stages, the detector set-up and the parallel read-out cabling are highlighted, b) Copper heat sinks for cable thermalisation through the cryostat temperature stages, c) A single ribbon cable containing 30 wires for three channels.}\label{FIG:cryogenics}
\end{figure}

Figure \ref{SUBFIG:ECHo_cryo} shows the open cryostat equipped with the cabling described in the following.
The wiring consist of 64 parallel SQUID read-out channels and 8 multi-purpose read-out channels for a total of 72 read-out channels that have been installed in the cryostat and exploited to operate two ECHo-1k detector chips in parallel during the ECHo-1k high statistics measurement campaign.

Two stacks of amplifier modules, each containing 36 read-out channels as described previously, are placed on the mixing chamber plate of the cryostat, as can be seen in figure \ref{SUBFIG:ECHo_cryo}. Altogether, 720 wires (10 wires per read-out channel) connect the amplifier stages to the room temperature electronics. 
The read-out wires are organised in twisted pairs and twisted triples interwoven in ribbons, each one containing the 30 wires required to read out three SQUID channels. The ribbons are stably woven with Nomex fibers\footnote{Manifactured by DuPont de Nemours GmbH, Hugenottenallee 175, 63263 Neu Isenburg, Germany.} resulting in $13 \, \mathrm{mm}$ wide, $0.7 \, \mathrm{mm}$ thick and $2.2 \, \mathrm{m}$ long flat ribbons\footnote{Produced by Tekdata Interconnections Limited, Innovation House, The Glades, Festival Way, Etruria, Stokeon-Trent, Staffordshire, ST1 5SQ, United Kingdom.}. 
The two ribbon ends are equipped with connectors: at the room temperature side a 24-pin LEMO connector\footnote{Connector EGG.3B.324.ZLL produced by LEMO, Chemin de Champs-Courbes 28, CH-1024 Ecublens, Switzerland.} interfaces a vacuum feedthrough\footnote{Coupler SGJ.3B.324.CLLPV produced by LEMO.} followed by the SQUID electronics. 
At the low temperature side a double row 30-pin connector\footnote{Connector type "SFM-115-01-S-D" produced by Samtec.} matches the connectors of the amplifier modules described in section \ref{SEC:SQUID_readout} (figure \ref{SUBFIG:ribbon}).
The wire material is optimised to reach a compromise between low resistance and low thermal conductance, to minimise the contribution to the heat input.
Alloy-30\footnote{Purchased from Isabellenhütte, Eibacher Weg 3-5, 35683 Dillenburg, Germany.}, an alloy of 2 \% nickel in copper, has been chosen. The wires have a diameter of $200 \, \mathrm{\upmu m}$ and a length of about $2.2 \, \mathrm{m}$. The round-trip resistance of a twisted pair of such wires is $6.7 \, \mathrm{\Omega}$ and drops only by about 20\% when cooled down. This is low enough that both the voltage noise of the warmer fraction of the wires and the current noise of the input amplifier of the SQUID electronics lead to contributions of the total noise which are marginal.

In order to dump any power that is dissipated in the wires and to reduce the heat load from room temperature to the mixing chamber plate, a good thermal coupling between the ribbons and the different temperature stages of the cryostat is necessary. In‐house manufactured OFC-copper heat sinks have been designed and installed (figure \ref{SUBFIG:heat_sinks}) \cite{Allgeier2017}. They consist of seven $46 \, \mathrm{mm} \times 50 \, \mathrm{mm} \times 5 \, \mathrm{mm}$ large copper plates screwed to each other. 
Six of them feature a milled channel which allows to place two ribbons side by side between each pair of plates, ensuring full surface contact and avoiding mechanical stress of the wires. To maximise the thermal coupling, vacuum grease\footnote{Apiezon N, M\&I Materials Ltd., Hibernia Way, Trafford Park, Manchester M32 0ZD, United Kingdom} is distributed on the contact area between ribbon and copper plates and the copper plates. The heat sinks at the two lowest temperature stages have been annealed at $800 \, ^{\circ} \mathrm{C}$ for 48 hours to remove hydrogen molecules.

The ribbons are connected at room temperature to aluminium boxes on top of the cryostat, where 24-pin vacuum feedthroughs connectors\footnote{Connector type SGJ.3B.324.CLLPV purchased from LEMO Elektronik GmbH, Hanns-Schwindt-Str. 6, 81829 Munich, Germany.} are used as interface between the atmospheric pressure outside and the vacuum inside the cryostat and provide the possibility to directly plug in the SQUID electronics used for the ECHo experiment\footnote{3XXF-1, Magnicon GmbH, Barkhausenweg 11, 22339 Hamburg, Germany.}. The SQUID electronics can be connected via LEMO cables to the SQUID connector box, a device which serves as power supply for the SQUID electronics and allows for tuning of the SQUID parameters via software.

\clearpage

\section{Data acquisition and processing}

All the parameters of the programmable two-stage SQUID electronics can be set by software\footnote{SQUID Viewer provided by Magnicon.}.
The output voltage of the linearised SQUID signal is transferred from the SQUID connector box to two ADC modules using coaxial cables equipped with NIM/Camac standard connectors\footnote{Produced by LEMO}. 
Each ADC module is based on a 16-channel digitiser card SIS3316\footnote{Produced by Struck Innovative Systeme GmbH, Harksheider Str. 102, 22399 Hamburg, Germany. \label{Struck}}.
The cards can either be read out via ethernet or via a SIS1100/3100\textsuperscript{\ref{Struck}} connection based on an optical gigabit link. 
If more than one module is used, their clocks can be synchronised, which is essential for coincidence measurements.
They feature a maximum sampling rate of $125 \, \mathrm{MHz}$ per channel and a resolution of $16 \,  \mathrm{bit}$. Supporting simultaneous acquisition of data and data transfer to the computer as well as individual asynchronous constant fraction trigger engines for each channel, the cards are suitable for high throughput and flexible read-out.
Oversampling up to a factor of 256 can be used to acquire larger time windows and to increase the effective resolution of the ADCs from about $13 \,  \mathrm{bit}$ to the full $16 \,  \mathrm{bit}$. At the same time a high timing precision on the fast clock is available to investigate coincidences.

An in-house developed C++ acquisition software \cite{Hengstler2017} allows to control hardware settings, data transfer, on-line signal processing and data storage. The software offers a graphical user interface with a virtual oscilloscope that allows to tune the SQUID parameters before starting the data acquisition and to perform noise measurements to further optimise the SQUID tuning or to characterise the detector properties. 
When the acquisition is started, for each triggered event the raw signal, typically containing 16384 samples, is saved on disk and relevant information (e.g. signal height, timestamp, polarity, ...) are determined and stored. Subsequently, an individual on-line fitting procedure based on a template fit takes place and the fit results are saved.
Additionally, for every triggered signal also a voltage trace of one or several temperature-sensitive channels can be acquired simultaneously to correct for signal height variations due to variations of the detector temperature.
Furthermore, the program allows for a software hold-off and a simple on-line rejection of signals, for example based on signal height and shape, is available.

\section{Characterisation of the read-out chain for ECHo}

The SQUID read-out channels which have been used during the first phase of the ECHo experiment have been individually characterised. The total read-out system consists of 64 two-stage SQUID read-out channels distributed in two separate set-ups, connected to two ECHo-1k detectors, respectively.

The front-end SQUIDs are characterised by an input coil inductance between $1.2 \, \mathrm{nH}$ and $1.6 \, \mathrm{nH}$, depending on the SQUID design. The typical value of the current swing of the front-end SQUIDs is about $5 \, \mathrm{\upmu A}$, matching the input current sensitivity of the amplifier SQUIDs of about $11.5 \, \mathrm{\upmu A / \Phi_0}$ and $8.4 \, \mathrm{\upmu A / \Phi_0}$, for the two set-ups, respectively.
The average mutual inductance of the front-end feedback coil is $42.4 \, \mathrm{\upmu A / \Phi_0}$ ($41.1 \, \mathrm{\upmu A / \Phi_0}$) with a standard deviation of $3.2 \, \mathrm{\upmu A / \Phi_0}$ ($1.6 \, \mathrm{\upmu A / \Phi_0}$) for the first (second) set-up, ensuring a homogeneous voltage response in all the channels of each set-up.
The amplification SQUIDs feature a peak-to-peak voltage swing of about $465 \, \mathrm{\upmu V}$ and $240 \, \mathrm{\upmu V}$ for the two set-ups, respectively.

The flux noise spectrum of all channels is mainly composed by two contributions. The first contribution is related to the thermodynamic properties of the detector, while the second one is given by the read-out noise. The first contribution cannot be reduced once the detector geometry and material as well as the operating temperature are fixed. The read-out contribution can be well described by two components. In the low frequency range the $1/f$ flux noise component is dominating, while in the high frequency range the white noise is prevailing.

\begin{figure}[h!] 
	\centering
	\includegraphics[width=0.6\textwidth]{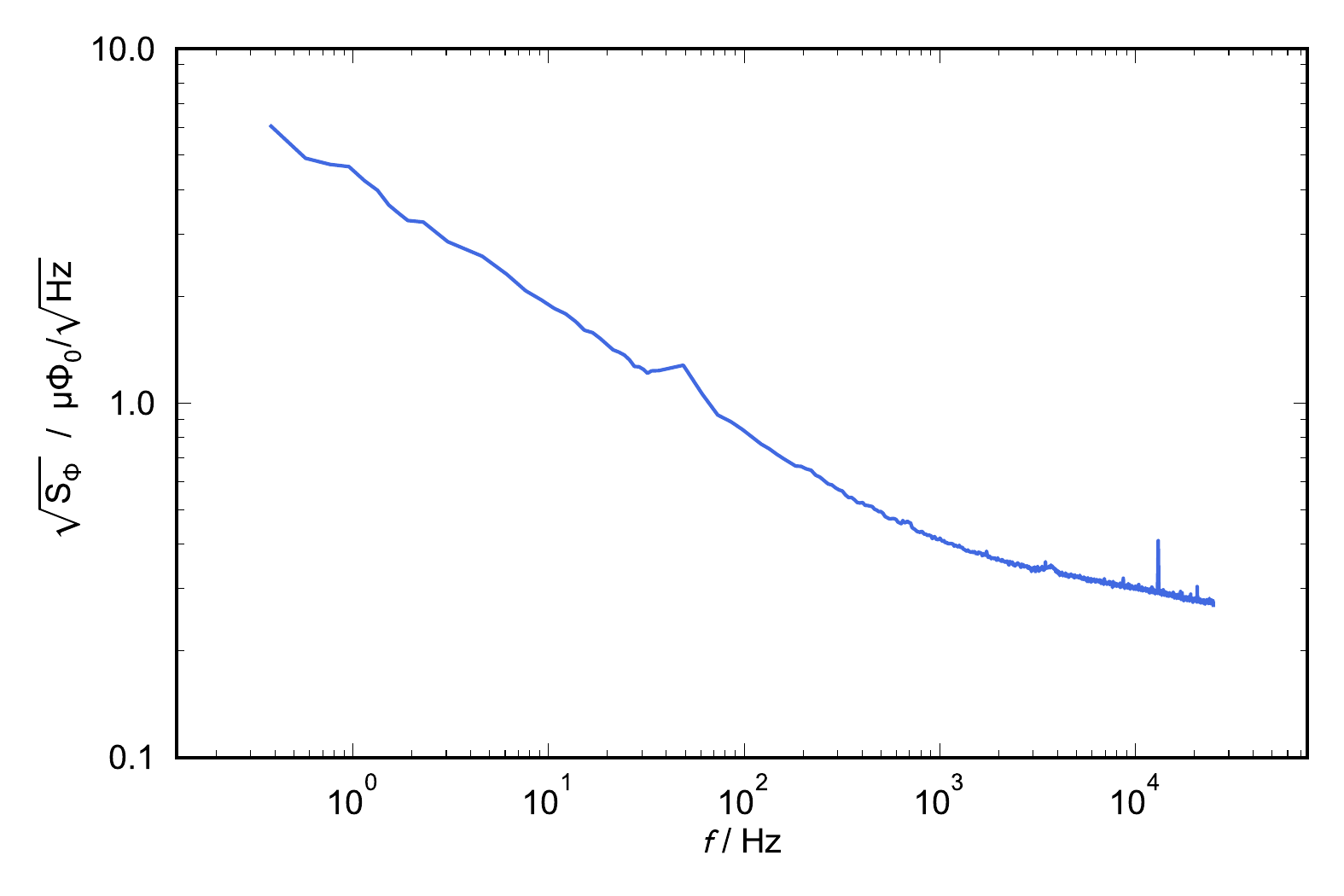}
	\caption{A noise measurement of a two-stage read-out channel with an MMC detector of the ECHo set-up connected. The measurement is performed in the low frequency range between $\mathrm{0.4 \, Hz}$ and $\mathrm{25 \, KHz}$.} 
	\label{FIG:noise}
\end{figure}

Figure \ref{FIG:noise} shows an example noise measurement of the ECHo read-out performed with an anti-aliasing filter with a cut-off of $100 \, \mathrm{KHz}$ and with no persistent current in the detector pick-up coil. The anti-aliasing filter affects the noise spectrum shown in figure \ref{FIG:noise} by causing a roll-off at frequencies above about $20 \, \mathrm{kHz}$.
Most of the operated two-stage channels show a white noise level of about $0.3 \, \mathrm{\upmu \Phi_0 / \sqrt{Hz}}$ and the a $1/f$ noise level at $1 \, \mathrm{Hz}$ of about $5.0 \, \mathrm{\upmu \Phi_0 / \sqrt{Hz}}$.

The two-stage SQUID read-out has been used for the ECHo-1k phase and the achieved noise performance is sufficient for this stage of the experiment. In fact, the planned energy resolution for the ECHo-1k phase is at least $10 \, \mathrm{eV} $FWHM and this requirement is fully satisfied. The average energy resolution achieved during the final ECHo-1k high-statistics measurement is about $5.5 \, \mathrm{eV}$ FWHM. For the coming second phase of the ECHo experiment, a microwave SQUID multiplexing read-out system is foreseen in order to scale up the number of pixels from about 100 to about 10000.

\section{Conclusions}

\begin{figure}[h!] 
	\centering
	\includegraphics[width=0.95\textwidth]{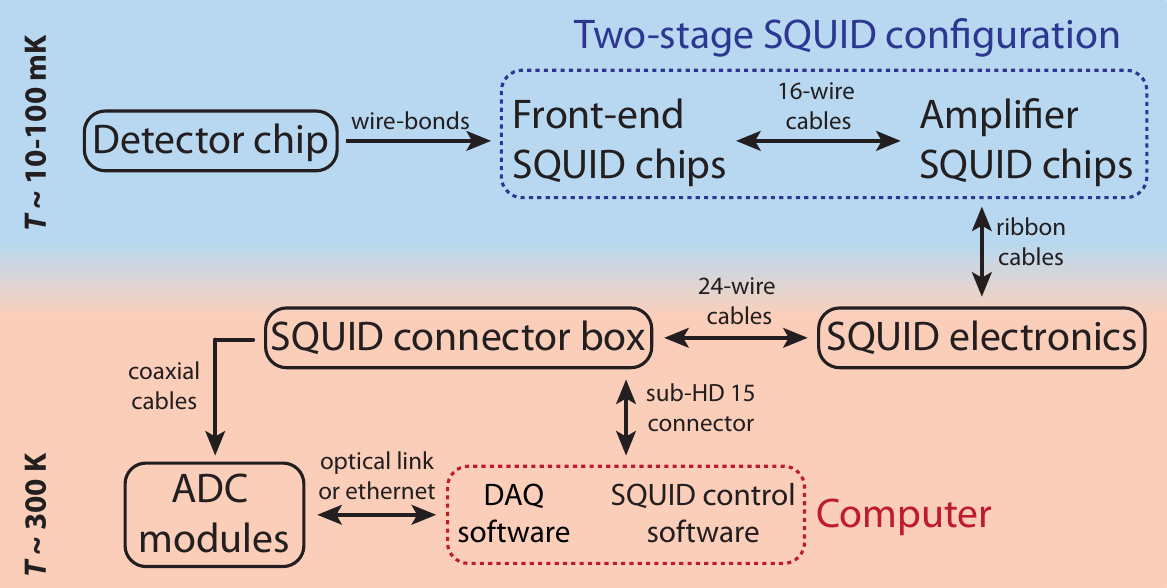}
	\caption{Schematic layout of the complete read-out chain for the ECHo experiment. The direction of the arrows refers to the data transfer (i.e. signal data and control signals) flow.} 
	\label{FIG:block_diagram}
\end{figure}

The typical read-out system for MMC arrays with 32 SQUID read-out channels includes the detector module, the amplifier module, the cabling from room temperature to the cryogenic platform and the data acquisition system. A standard layout has been defined for each component of this read-out chain, except for the detector module, which has to be optimised for the specific detector chip design and has to fulfil the requirements defined by the particular application. The resulting read-out system is compatible with different detector modules and, due to the modular approach, offers high flexibility. 
In order to illustrate the different components of the 32 channels MMC read-out, we have presented the system developed for the first phase of the ECHo experiment, ECHo-1k. A read-out system for the operation of two ECHo-1k modules has been set up and is schematically depicted in figure \ref{FIG:block_diagram}. 
The detector chip consists of 36 double meander channels including two non-gradiometric channels for temperature monitoring and two test channels. For the experiment, 32 channels for each of the two chips were connected to the read-out chain. 
The first read-out stage of each channel consists of single-stage dc-SQUIDs, specifically designed to match the detector requirements, directly glued onto the detector holder and wire-bonded to the detector. The second stage consists of a 16 dc-SQUID series array, provides amplification at low temperature and is placed in separate modules shielded by tin plated copper casings. Both the detector module and the amplifier modules are screwed to the mixing chamber plate of a dry dilution refrigerator. 

The cryostat is equipped with two sets of read-out cables, each set consisting of 36 read-out channels, connecting the detector module at the mixing chamber stage at a temperature of about $20 \, \mathrm{mK}$ to the room temperature electronics. 
The wire material and size have been optimised to guarantee minimal contribution to the heat load. Custom-built copper heat sinks are installed at each temperature stage to ensure reliable cable thermalisation. The room temperature side of the cables is interfaced with the vacuum inside the cryostat trough aluminium casings equipped with vacuum feedthroughs. The SQUID electronics are plugged to the vacuum feedthroughs on one side and to the SQUID connector box on the other side. The signal is then transferred to two ADC modules based on $125 \, \mathrm{MHz}$ sampling rate digitiser cards supporting high signal throughput and individual asynchronous trigger for all channels. Hardware settings, data acquisition and data storage are controlled by software, which offers also on-line signal processing and on-line fitting.

The same read-out scheme has been installed in other dilution refrigerators, as the one to be integrated at the Cryogenic Storage Ring (CSR) \cite{CSR} for the operation of the MOCCA detector \cite{Gamer2016} or the cryostats operated in collaboration with the GSI and the HI Jena for the study of highly charged ions with different detectors of the maXs family \cite{Hengstler2015}. 
The availability of standard parts for the read-out of 32-channels MMC arrays allows not only for well characterised properties of each component, but also for a fast exchange of damaged parts. 

\section*{Aknowledgments}

Part of this research has been performed in the framework of the DFG Research Unit FOR 2202 (funding under GA 2219/2-2, EN 299/7-2, EN 299/8-1). 
We acknowledge the support of BMBF under the contract 05P19VHFA1.
F. Mantegazzini, A. Barth, D. Schulz, D. Unger, C. Velte and M. Wegner acknowledge the support by the Research Training Group HighRR (GRK 2058) funded through the Deutsche Forschungsgemeinschaft, DFG.
We acknowledge the support of the cleanroom team of the Kirchhoff-Institute for Physics, Heidelberg University. 


\end{document}